# Uncovering Electronic Exchange Behavior: Exploring Insights from Simple Models

Rezaei, Mahnaz[1]; Abouie, Jahanfar,[2,*] Nazari, Fariba[1,3,*]

[1]Department of Chemistry, Institute for Advanced Studies in Basic Sciences, Zanjan 45137-66731, Iran

[2]Department of Physics, Institute for Advanced Studies in Basic Sciences, Zanjan 45137-66731, Iran

[3]Center of Climate Change and Global Warming, Institute for Advanced Studies in Basic Sciences, Zanjan 45137-66731, Iran

**ABSTRACT:**

Exchange couplings are fundamental to our understanding of many physical phenomena in condensed matter physics and material science. Model systems provide a controlled environment to investigate such phenomena, effectively. In this study, we employ first-principle calculations based on density functional theory and Green's function (GF) method to explore the impact of chemical structure on the sign and magnitude of exchange coupling, systematically. By designing model systems with bcc-Fe bulk doped with nonmagnetic X= (H, B, C, N, O, and F) atoms, we examine the effects of different ligands on the behavior of Fe-Fe exchange coupling, and demonstrate that the chemical environment surrounding the metal atom significantly influences the Fe-Fe exchange coupling. Our results highlight the tunability of exchange coupling based on Fe-dopant bond length(s), where the nature of ligand atoms and their electron correlation play a crucial role. This work illuminates the complex relationship between structure, and magnetism in magnetic materials, providing insights into the development of high-performance magnetic materials.     *Corresponding authors: jahan@iasbs.ac.ir, nazari@iasbs.ac.ir,

# 1. INTRODUCTION

Model systems provide a controlled environment for researchers to investigate fundamental principles underlying complex phenomena, effectively. These systems are employed across diverse scientific disciplines, serving as tools to disentangle complicated mechanisms such as alloy compounds, test hypotheses, and generate insights. The insights gained from model systems often have practical implications, driving translational applications and innovations.

Magnetism, one of the fundamental disciplines where theoretical investigations are characterized by significant complexity, can be challenging due to the interplay of various factors such as the type of magnetic atoms, separation distance, and chemical environment capturing the behavior of magnetic materials accurately; hence, the utilization of a model system can be beneficial in this regard.

Today, controlling the properties of magnetic materials has received considerable attention due to the common fundamental interest and technological demand for faster and more efficient magnetic storage [1-3]. The very promising pathway for the control of magnetic properties of materials is the engineering of the exchange coupling between microscopic magnetic moments. The magnetic properties of materials depend on the magnitude and sign of the exchange coupling, and in fact, the interplay of the exchange coupling and thermal fluctuations determines the magnetic behavior of a material at a finite temperature. Effectively tuning of exchange couplings necessitates a comprehensive understanding of how the structural characteristics including chemical environment and lattice parameter of magnetic materials delicately influence their magnetic properties. Undoubtedly, exploring rational strategies to tune and control the structure of magnetic materials and their magnetic properties are crucial. Such efforts are essential for determining the relationship between the structural characteristics and magnetic properties of a material, and consequently for development of high-performance magnetic materials.

Recently, it has been experimentally [4] and theoretically [4-7] demonstrated that the exchange coupling between two magnetic centers in single-molecule magnets can be controlled by two parameters; the bridging angle and the distance of magnetic centers. Also, it has been experimentally demonstrated that the exchange coupling between two dimeric manganese (III) compounds can be controlled through the bridging ligands [8]. Despite significant efforts in the field of exchange coupling engineering for magnetic systems, fully understanding, modeling, and controlling magnetic properties through exchange coupling remains challenging, particularly in periodic materials. For instance, the literature does not sufficiently address how metal-ligand distances and the types of ligands influence exchange coupling, highlighting a critical gap in knowledge that requires further investigation. Such research is essential for strategically manipulating these couplings and will provide valuable insights for the development of high-performance magnetic materials. To address this gap, we employ first-principles calculations based on density functional theory (DFT) to systematically explore the effects of magneto-structural correlations on both the sign and magnitude of exchange coupling . To this end, we design model systems by considering bcc-Fe bulk doped with nonmagnetic (X = H, F, N, C, B, and O) atoms, and investigate the effects of these nonmagnetic impurities on the behavior of Fe-Fe exchange coupling. Our study reveals that the chemical environment surrounding the metal atom plays a critical role in the sign and magnitude of Fe-Fe exchange coupling. In particular, we demonstrate that the exchange coupling strength can be changed by tuning the Fe-X separation distance. Actually, when X atoms are bonded with a magnetic Fe atom through coordinate covalent (dative) bonds, the Fe-X-Fe bond length determines the indirect overlapping of the two Fe atoms residing at the end of Fe-X-Fe bond, and thus plays an important role in the magnitude of the Fe-Fe exchange coupling.

This demonstrates the sensitivity of exchange coupling to the metal−ligand distance(s).

The rest of this paper is organized as follows. In Section 2 we present computational details. In this section we employ isotropic Heisenberg spin model to obtain the exchange coupling between Fe atoms. In Section 3 our main results are discussed, and finally in Section 4 we summarize our results.

## 2. COMPUTATIONAL DETAILS

In our study, spin-polarized (SP) electronic structure calculations were performed using the density functional theory (DFT) framework, as implemented in the Quantum Espresso code [9]. We utilize the Perdew-Burke-Ernzerhof revised for solids (PBEsol) [10] exchange-correlation functional within the generalized gradient approximation [11] The ion-electron coupling is treated by the projector augmented wave method [12]. For sampling of the Brillouin zone, Γ-centered Monkhorst-Pack [13] $k$-point meshes of $6 \times 6 \times 6$ are used for all systems, and an energy cutoff of 1224.51 eV is adopted for the plane-wave basis. We have calculated the Hubbard parameter (U=5 eV) for the bcc-Fe bulk using **hp.x**. It is based on density functional perturbation theory. We implemented this parameter across other systems.

The magnetic moment of the Fe atom ($\mu$) residing at site $i$ is obtained by utilizing the following formula [14, 15]:

$$\mu_i = \int_0^{r_m} \left[ n_i^\uparrow(r) - n_i^\downarrow(r) \right] d^3r, \tag{1}$$

where $n_i^\uparrow(r)$ and $n_i^\downarrow(r)$ are the spin-up and spin-down electron densities, respectively. The integrals in Eq. (1) are computed locally around the Fe atom ($r_m$). The rationale for setting $r_m$ as a fraction of the minimum interatomic distance ensures that the integration spheres for each atom do not overlap [9]. The magnetic moment of the Fe atom can also be given in terms of the projected density of states on the Fe atom especially $d$ orbitals because magnetism originates from the $d$ orbitals of the Fe atom as $\mu = N_d^\uparrow - N_d^\downarrow$, where $N_d^\sigma = \int g_d^\sigma(E)\, dE$. Here, $g_d^\sigma(E)$ is spin-$\sigma = \uparrow, \downarrow$ electrons density of states of the $d$ orbitals.

A deep fundamental knowledge of the relationships between structure and electronic properties, which are difficult to access from experiments, allows us to uncover how the chemical environment influences exchange couplings. By using these data, we investigate the exchange coupling between magnetic moments on all the model systems.

One of the most common and successful microscopic models for describing magnetic properties of systems with inversion symmetry is the Heisenberg model [16] with localized spins:

$$H = \sum_{i \neq j} J_{ij} S_i \cdot S_j , \qquad (2)$$

where $J_{ij}$ is the exchange coupling of the spins residing at sites $i$ ($S_i$) and $j$ ($S_j$). The relation between the magnetic moment ($\mu$) and the spin (S) of Fe atoms are expressed as $\mu = g\, \mu_B (S(S+1))^{1/2}$ where $g$ called $g$-factor.

There are a few methods to estimate $J_{ij}$ within SP-DFT. The most straightforward and widely used approach to determine $J_{ij}$ is to calculate the total energies of the P + 1 magnetic configurations, where P represents the number of distinct exchange couplings [17-19]. However, this method presents several notable drawbacks: (a) it requires generating a substantial number of distinct magnetic configurations for complex systems; (b) all configurations must maintain the same magnetic moments, which is particularly important for materials close to the itinerant regime; and (c) the exchange coupling obtained is a number, making it difficult to distinguish which orbitals significantly influence the exchange coupling behavior and which mechanisms, such as direct or indirect exchange, are responsible for it [20]. To address these limitations, the Green's function (GF) method is employed [21-23]. In this approach, analytical expressions have been derived for the changes in total energy due to small spin rotations via the magnetic force theorem [21, 24, 25], utilizing SP-DFT and the Heisenberg model. Within this approach we can obtain both the total exchange couplings from a single magnetic configuration, and orbitals contributions to the total exchange coupling. The approach developed by Dm. M. Korotin et al. [20] has been adapted for plane-wave methods through the formalism of Wannier functions (WFs) [23, 26]. We utilized this methodology, where the transformation of plane-wave eigenstates to WFs was performed using the *wannier_ham.x* utility available with the Quantum Espresso code.

All the above-mentioned model systems, i.e., the bcc-Fe bulk doped with nonmagnetic X atoms, possess inversion symmetry. Moreover, the nonmagnetic X atoms, serving as ligands, are light elements with very weak spin-orbit coupling. In this respect, the Fe-Fe exchange couplings can be incorporated in a generalized Heisenberg Hamiltonian:

$$H = \sum_{i \neq j} J_{ij} e_i e_j \tag{3}$$

where $e_i$ and $e_j$ refers to the unit spin vectors of atoms residing at sites $i$ and $j$. A negative (positive) sign of $J_{ij}$ indicates a ferromagnetic (antiferromagnetic) coupling. In order to compute the exchange coupling $J_{ij}$, all orbitals contribution should be taken into account [27]. The couplings between different $d$ orbitals of two Fe atoms make a $5 \times 5$ matrix given by:

$$\mathcal{J}_{ij} = \begin{pmatrix} d_{z^2} d_{z^2} & d_{z^2} d_{xz} & d_{z^2} d_{yz} & d_{z^2} d_{x^2-y^2} & d_{z^2} d_{xy} \\ d_{xz} d_{z^2} & d_{xz} d_{xz} & d_{xz} d_{yz} & d_{xz} d_{x^2-y^2} & d_{xz} d_{xy} \\ d_{yz} d_{z^2} & d_{yz} d_{xz} & d_{yz} d_{yz} & d_{yz} d_{x^2-y^2} & d_{yz} d_{xy} \\ d_{x^2-y^2} d_{z^2} & d_{x^2-y^2} d_{xz} & d_{x^2-y^2} d_{yz} & d_{x^2-y^2} d_{x^2-y^2} & d_{x^2-y^2} d_{xy} \\ d_{xy} d_{z^2} & d_{xy} d_{xz} & d_{xy} d_{yz} & d_{xy} d_{x^2-y^2} & d_{xy} d_{xy} \end{pmatrix}, \tag{4}$$

where the conventional value of $J_{ij}$ corresponds to the sum of all the above matrix elements:

$J_{ij} = \sum_{mm'} \mathcal{J}_{ij}^{mm'}$ with $m$ and $m'$ being the five $d$ orbitals.

For computing the Fe-Fe exchange coupling, we employed Green's function formalism [21-23]. Within this formalism the matrix elements of the exchange coupling $\mathcal{J}_{ij}$ in Eq. (4) is expressed as:

$$\mathcal{J}_{ij}^{mm'} = -\frac{1}{2\pi}\text{Im}\int_{-\infty}^{E_F} d\varepsilon \sum_{nn'n''}\left(\Delta_i^{mn} G_{ij,\downarrow}^{nn'}(\varepsilon) \Delta_j^{n'n''} G_{ji,\uparrow}^{n''m'}(\varepsilon)\right), \qquad (5)$$

where the sums on $n, n', n''$ run over different orbitals at sites $i$ and $j$. Here, $E_F$ is Fermi energy, $G_{ji,\uparrow}^{mm'}\left(G_{ij,\downarrow}^{mm'}\right)$ is the real-space intersite Green's function for spin-up (-down) electrons residing in orbitals $m$ and $m'$, and $\Delta_i^{mm'}$ is the local exchange splitting.

The convergence of the total exchange coupling ($J_{ij}$) and the magnetic moment with respect to the number of $k$-points is evaluated using progressively finer $k$-point meshes. Specifically, we tested $k$-point meshes of 16 × 16 × 16, 40 × 40 × 40, and 50 × 50 × 50 to determine their accuracy. Our results indicate that a 16 × 16 × 16 mesh provides adequate convergence for these parameters. We have used 16 × 16 × 16 $k$-points mesh for all the systems.

## 3. RESULTS AND DISCUSSION

The aim for choosing the bcc-Fe bulk as our preferred prototype material for the ab initio study of magnetic properties is twofold: (i) its simple structure, and (ii) the abundance of both experimental and theoretical data available to evaluate our computational results.

The ligand atoms around a Fe atom generate a crystal-field which interacts with electrons localized in $d$ orbitals of the corresponding Fe atom, and cause energy-level-splitting of the $d$ orbitals. According to crystal-field theory, in a bcc crystal lattice the five $d$ orbitals, $d_{xy}, d_{yz}, d_{xz}, d_{x^2-y^2}$, and $d_{z^2}$, get hybridized and form $t_{2g}$ ($d_{xy}$, $d_{yz}$, and $d_{xz}$) and $e_g$ ($d_{x^2-y^2}$, and $d_{z^2}$) states [28].

Various factors such as ligands type around the Fe atom, Fe-ligand distance, and crystal symmetry affect crystal-field and influence the magnetic and electronic properties [29, 30]. By understanding and manipulating the crystal-field, researchers can tailor the properties of systems for specific applications.

To this end, we designed the following model systems using the bcc-Fe bulk:

I. The bcc-Fe bulk with the equilibrium lattice constant $a$ ($a$=2.86 Å) and twice its value $2a$. These model systems are respectively labeled as 1$a$Fe and 2$a$Fe (see Fig. 1a, c).

II. The bcc-Fe bulk with the equilibrium lattice constant $a$ doped with X (X = H, F, N, C, B, and O) atoms. This model system is labeled as 1$a$FeX, where the X dopant is placed right at the middle of the two Fe atoms ($L_m$) on the sides of the cube. Here, the Fe-X bond length is $L_m = 1.43$ Å (see Fig. 1b).

III. The bcc-Fe bulk with lattice constant $2a$ doped with X atoms (X = H, F, N, C, B, and O). X atoms can be located at different positions between two Fe1 atoms. These systems are labeled as 2$a$FeXL$_i$, where $L_i$ with i=1, 2,…,6 is the Fe-X separation length (see Fig. 1d and Table 1). In each 2$a$FeXL$_i$ system one of the $L_i$s is the Fe-X bond length measured in available real systems. In particular, $L_1$ is the Fe-H bond length in a real Fe-H system, $L_2$ is the Fe-F bond length in Fe-F system, and $L_3$, $L_4$, $L_5$, and $L_6$, are the Fe-N, Fe-C, Fe-B, and Fe-O bond lengths in the corresponding systems, respectively. In addition, we have also considered 2$a$FeXL$_m$ systems in which the X atoms are exactly located at the middle of two Fe1 atoms.

**3.1. The 1*a*Fe System.** We begin by computing the magnetic moment of the Fe atoms and studying the behavior of the Fe-Fe exchange coupling in the 1*a*Fe system.

Our results show that the magnetic moment of Fe atoms in the 1*a*Fe system is $2.7\mu_B$. As we will demonstrate in the next sections the magnetic moment of the Fe atoms strongly depends on the lattice constant, and the increase of the lattice constant from $a$ to $2a$, significantly enhances the magnetic moment of the Fe atoms, in amount.

We have also computed the Fe-Fe exchange coupling in the 1*a*Fe system and found that the total exchange couplings values along the diagonal and the edge directions are $J_1$ = -26.734 meV and $J_2$= -18.318 meV, respectively. Our results are in agreement with those of reported in Ref. [31]. We have computed the exchange coupling of the Fe atoms with their first to seventh nearest neighboring atoms along each of the diagonal and the edge directions. As we have illustrated in Fig. 2, similar to the Ruderman-Kittel-Kasuya-Yosida (RKKY) couplings [32-36], both the $J_1$ and $J_2$ couplings decay oscillatory by increasing the interatomic distance of Fe atoms. In order to understand the reason of such an oscillatory decay of the exchange couplings, we have also computed the exchange coupling of different orbitals, contributions, i.e., the exchange couplings of electrons occupying $e_g - e_g$ couplings orbitals, $t_{2g} - t_{2g}$ couplings orbitals, and $e_g - t_{2g}$ couplings orbitals, separately. In Fig. 2a, we have plotted $J_1$ for the different orbital couplings as a function of Fe-Fe interatomic distance. The oscillatory behavior is seen in all the orbital couplings; However, the primary dominant contribution originates from the $t_{2g}$ states.

To analyze this, we have also plotted in Fig. 3 the projected density of states (PDOS) of the $d$ orbitals for the 1*a*Fe system. As seen, the density of states (DOS) of the $t_{2g}$ orbitals dominate near the Fermi level, and is much broader (delocalized) than the DOS of the $e_g$ orbitals. This observation underscores the significance of the $t_{2g}$ orbitals in determining the electronic structure and properties of the system near the Fermi level. This is in alignment with previous findings, as noted in Ref. [37], underscores the crucial role played by the $t_{2g}$ orbitals in the electronic structure at the Fermi level. Also, one can observe that the $t_{2g} - e_g$ coupling along the edge direction has negligible contribution (see Fig. 2b). In principle, this term is likely to be forbidden by symmetry. Therefore, the mixed $t_{2g} - e_g$ coupling has no contribution in the $J_2$ coupling [38].

**3.2. The 1aFeX system.** As we mentioned before, the X= H, F, N, C, B, and O dopant is placed right at the middle of the two Fe atoms ($L_m$) on the sides of the cube (interstitial site). Previous studies have indicated that in the bcc-Fe bulk, small foreign atoms such as C and N tend to occupy this site [39, 40]. Doping the bcc-Fe bulk with X= H, F, N, C, B, and O atoms reduces the magnetic moment of the Fe atoms (see Fig. 4). The most significant reduction is observed for Fe1, particularly in the systems with X=F and B, where the magnetic moment of the Fe1 atoms is almost zero. As we discussed in Eq. (1), the reduction in the magnetic moment is due to the changes of the PDOS of the Fe atoms near Fermi level, when X atoms are doped to the system. To illustrate this, we have also plotted in Fig. 5 the PDOS corresponding to of different orbitals for 1aFeX systems. As the most significant reduction is observed for the Fe1 atoms, we have focused on the PDOS of Fe1 atoms. As seen from Figs. 5b and 5e, the PDOS for the spin-up and spin-down *d* electrons of Fe1 atoms and the PDOS for the *p* orbitals of the X atoms in the 1aFeF and 1aFeB systems are almost the same. Moreover, there is a significant decrease in the PDOS for spin-up and spin-down *d* electrons of the Fe1 atom at the Fermi level in the 1aFeX systems with X=H, N, C, and O compared to the pristine 1aFe system (see Fig. 5a, 5c, 5d, and 5f).

The change in the magnitude of PDOS for spin-up and spin-down $d$ electrons of Fe atoms can be explained within the frame work of the concept of the electron-transfer from X atoms to the $d$ orbitals of the Fe atoms [41, 42]. The electron-transfer for systems with X=F and B is greater, leading to a sharp decrease in the spin density and ultimately in the magnetic moment of Fe1 atoms.

Doping the 1$a$Fe system with the above mentioned nonmagnetic atoms also significantly modifies the strength of the orbital couplings, as depicted in Fig. 6. As our results clearly show, the $e_g - e_g$ contribution in both the $J_1$ and $J_2$ couplings is significantly reduced by doping. In this context, it is evident that the axes of the coordinate system ($x$, y, $z$) coincide with the axes of the cubic crystal system ($a$, $b$, c) and the $e_g$ axial orbitals are oriented along the $a$, $b$ and $c$ axes. In addition, X atoms can also locate along the cubic crystal axes directions. Therefore, the axial orbitals of the desired X-doped structures have the most effectiveness and influence. As a result, the orbital coupling associated with these orbitals is significantly influenced by doping. Another important observation is that when X atom locates at $L_m$ the symmetry does not change, and consequently, similar to the 1$a$Fe system, the mixed $t_{2g} - e_g$ coupling has vanishing contribution in $J_2$ exchange coupling.

We have also investigated different orbital contributions to the $J_1$ and $J_2$ couplings as a function of Fe-Fe interatomic distance for the 1$a$FeX systems. As shown in Fig. 7, all types of orbital couplings contributions in the $J_2$ ($J_1$) exchange coupling are suppressed for the systems with X=C, F, B (X=F, B) dopants (see Fig. 7c, 7d, 7h, 7i, and 7k). However, in the systems doped with X= H, N, and O, the contribution of orbital couplings are nonzero for nearest neighbors (NNs) in both the diagonal and the edge directions, but significantly decrease by increasing the Fe-Fe interatomic distance, and falls down toward zero for beyond the second-neighbor of Fe atoms (see Fig. 7a, 7b, 7e, 7f, 7l, and 7m). Similarly, in system doped with X = C, the orbital coupling contributions are nonzero for NNs in the diagonal direction, but they also diminish rapidly with increasing Fe-Fe distance, falling to near zero beyond the second-neighbor Fe atoms (Fig. 7g).

The short-range exchange couplings between the Fe atoms in the 1$a$FeX systems are consistent with the reduction in the PDOS of the Fe1 atoms near the Fermi level. As we have already illustrated in Fig. 3, the delocalized states of the Fe atoms in the 1aFe system originate from the $t_{2g}$ orbitals which are pivotal in shaping the behavior of the RKKY couplings. Actually, the decrease in PDOS of the Fe1 atoms near the Fermi level in the 1aFeX systems (Fig. 5) results in a reduction in the strength and the amplitude of the $J_1$ and $J_2$ exchange couplings. Consequently, this results in the previously discussed short-range behavior of exchange couplings in the 1aFeX systems

**3.3. The Magnetic Moment and The Exchange Coupling in The 2$a$Fe System.** In this section, we study the effects of lattice constant increase on the behavior of exchange coupling and magnetic moment. We have computed the magnetic moment of the Fe atoms in both the pristine and doped 2$a$Fe systems. Our results show that an increase in the lattice constant leads to an enhancement of the magnetic moment of the Fe atoms. In particular, our results show that, by increasing the lattice constant from $a$ to $2a$ the magnetic moment of the Fe atoms increases to $4\mu_B$. To understand the reason behind such an enhancement, we plot the PDOS of the 2$a$Fe system (see Fig. 8), and compare it with the PDOS of the 1$a$Fe system. As seen from Figs. 3 and 8, increasing the lattice constant induces changes in the intensity and energy range of the PDOS, where two of them are much more pronounced: the first one is that the intensity increase of the DOS of both the $e_g$ and $t_{2g}$ orbitals, but the intensity increase in the PDOS of spin-up electrons is greater than those of spin-down electrons. The second one is that in contrast to $1a$Fe system, in $2a$Fe system the $t_{2g}$ and $e_g$ orbitals are localized in distinct and narrow energy regions without overlapping. These changes increase the magnetic moment of Fe atoms, as predicted by Eq. (1). We have also examined the variation of orbital couplings along the diagonal and the edge directions as a function of the Fe-Fe interatomic distance. As shown in Fig. 9, the exchange couplings decrease sharply by increasing of the lattice constant from $a$ to $2a$. The localization of $t_{2g}$ and $e_g$ states near the Fermi level in the $2a$Fe system, as depicted in Fig. 8, indicates that RKKY is not the dominant mechanism in this system. Direct overlaps between the atomic orbitals of Fe atoms specify the strength of Fe-Fe exchange coupling, however they are absent here due to an increased lattice constant. According to Fig. 8, the overlaps between atomic orbitals have been decreased, dramatically, consequently the $J_1$ and $J_2$ exchange couplings decrease, significantly.

**3.4. $2a$FeXL$_i$ Systems.** To investigate the effects of doping in the $2a$Fe system, we consider different cases for Fe-X separation length, as mentioned in Section 3, and presented in Table 1. As shown in Fig. 10, doping the $2a$Fe system with X=H, F, N, C, B, and O atoms results in a reduction of the magnetic moment of Fe1 and Fe2 atoms, compared to the pristine $2a$Fe system. The magnetic moment reduction is more pronounced for the Fe1 atoms in the $2a$FeXL$_i$ systems with X=C and B, because of more electron-transfer from X to Fe atoms in these systems.

We have also examined the behavior of the $J_1$ and $J_2$ exchange couplings. According to our results (not shown) X doping has more effects on the exchange coupling $J_2$ compared to $J_1$. In the following we will focus on the behavior of the $J_2$ coupling in the $2a$XL$_i$ systems.

In Fig. 11, we have illustrated the plot of $J_2$ exchange coupling as a function of $L_i$ for different $2a$XL$_i$ systems. By comparing the trend of the exchange coupling $J_2$ with respect to the distance presented in Figs. 11 and 9, one can see that doping the 2$a$Fe system with X atoms dramatically enhances the $J_2$ exchange coupling. Such an increase is due to the formation of an indirect superexchange coupling [43, 44] between Fe atoms, in consequence of the overlapping of $p$ orbitals of X atoms with $d$ orbitals of the Fe atoms.

We have also illustrated in Fig. 11, the contribution of different orbitals to the above mentioned superexchange coupling. As observed, nearly all orbitals contribute to the determination of the superexchange coupling value, $J_2$.

A more detailed investigations of the $J_2$ superexchange coupling show that each of the doped systems exhibits a maximum value ($J_2^{max}$) at a specific Fe-X separation distance (see Fig. 11). For example, for the 2$a$FeCL$_i$ system (2$a$FeFL$_i$ and 2$a$FeBL$_i$ systems) the $J_2^{max}$ is observed when the Fe-C (Fe-F and Fe-B) separation length is $L_3$ ($L_4$). As we presented in Table 1, $L_4 - L_3 = 0.011$ Å, therefore a very small change in the Fe-X separation length leads to significant changes in $J_2$. To show this, the Fe-X separation lengths where the superexchange $J_2$ is maximum are scanned for all model systems using a step size of 0.011 Å. The results are shown in Table S1. This observation confirms that the $J_2$ exchange coupling is highly sensitive to small variations in the Fe-X separation length. It is evident that the 2$a$FeXL$_s$ systems exhibit a maximum value ($J_2^{max}$) at a specific Fe-X separation length. The $J_2^{max}$ for each of the 2$a$FeXL$_s$ systems is reported in Fig. 12a.

It is well known that the energy difference of the $d$ and $p$ orbitals of Fe1 and X atoms, their overlaps, and $d$-$p$ hopping integral are major factors that determine the strength of the superexchange couplings [45, 46]. In Fig. 12b we have illustrated the energy difference of the $d$ and $p$ orbitals ($\Delta E^{\downarrow}_{e_g, p_{x(y,z)}}$, $\Delta E^{\downarrow}_{t_{2g}, p_{x(y,z)}}$) for different 2$a$FeXL$_s$ systems with $J_2^{max}$. As seen the $\Delta E^{\downarrow}_{t_{2g}, p_{x(y,z)}}$ for all the 2$a$FeXL$_s$ systems with $J_2^{max}$ are lower than the $\Delta E^{\downarrow}_{e_g, p_{x(y,z)}}$. Besides, to have an estimate for the *d-p* hopping integral, we have compared in Fig. 13, the integrated area of the unoccupied $e_g$ and $t_{2g}$ states of the Fe atoms across different energy intervals for the $2a\text{FeXL}_s$ systems with $J_2^{\max}$.

As clearly observed, in the $2a\text{FeXL}_s$ systems with X=F, C, and B, the unoccupied $t_{2g}$ states of Fe atoms are more delocalized and participate in hybridizations with the *p* states of the X atoms (see Fig. 13b, 13d, and 13e).

In addition, due to the smaller energy difference between $t_{2g}$ and *p* orbitals, it is easier for an electron to jump from one orbital to another, demonstrating the benefit of the $t_{2g}$-*p* hopping integral and the strong coupling of the $t_{2g} - t_{2g}$ and $e_g - t_{2g}$ in the $2a\text{FeXL}_s$ with X=F, C, and N and the $2a\text{FeBL}_s$ systems, respectively (see Fig. 12a).

In the case of the $2a\text{FeOL}_s$ system, the integrated area of the unoccupied $t_{2g}$ states across different energy intervals is localized compared to the integrated area of the unoccupied $e_g$ states, but they exhibit strong coupling (Fig. 13f). This is due to the fact that in the case of $2a\text{FeOL}_s$ system, the integrated area of the unoccupied $t_{2g}$ states close to the Fermi level is greater than the $e_g$ states. The presence of a more integrated area of the unoccupied $t_{2g}$ states near the Fermi level suggests a significant contribution to the electronic structure and less difference in energy between $t_{2g}$ and *p* orbitals, increases the probability of the electron-transfer from O atom as a bridging to Fe atoms. This ultimately leads to strengthening the $t_{2g} - t_{2g}$ coupling.

For the $2a\text{FeHL}_s$ system, although the energy difference between the $t_{2g}$ and *s* orbitals is lower than the $e_g$ and *s* orbitals, the $e_g - e_g$ couplings make a significant contribution. As observed in Fig. 13a, the integrated area of the unoccupied $e_g$ states are delocalized compared to $t_{2g}$ states in different energy intervals. Additionally, due to the spherical symmetry of the $s$ orbital of the H atom, its overlap with the $e_g$ orbitals are non-zero, emphasizing the dominant contribution of $e_g - e_g$ couplings.

An evident competition for $J_2^{max}$ is observed among all $2a\text{FeXL}_s$ systems (Fig. 12a). It is evident that the $2a\text{FeHL}_s$ ($2a\text{FeFL}_s$) system exhibits the highest (lowest) $J_2^{max}$ among the $2a\text{FeXL}_s$ systems (see Fig. 12a).

Fig. 14 clearly shows that the intensity of the integrated area of the unoccupied $d$ states for $2a\text{FeXL}_s$ systems with X=H, N, C, and O as the bridge has a similar order, while the delocalization of the integrated area of the unoccupied $d$ states for the H bridge is greater than other model systems. This has caused the $J_2^{max}$ to be higher for the $2a\text{FeXL}_s$ system with the H bridge than others. Also, the results confirm that electron correlation is a crucial factor in determining the exchange coupling.

Specifically, the H bridge with only one electron, as a bridge without electron correlation, exhibits the highest superexchange coupling while the F bridge, with the highest electron correlation, displays the least exchange coupling among all X atoms at a particular distance.

## 4. CONCLUSIONS

Controlling the properties of magnetic materials is increasingly important for advancing faster and more efficient magnetic storage technologies. The very promising pathway for the control is engineering the exchange coupling between microscopic magnetic moments. This requires a deep understanding of how structural characteristics, such as chemical environment and lattice parameters, influence magnetic properties.

To this end, we designed model systems based on bcc-Fe bulk with both the equilibrium lattice constant $a$ and twice its value $2a$. Then, we introduced nonmagnetic dopants (X = H, F, N, C, B, and O) and examined how these impurities affect the Fe-Fe exchange coupling behavior. Our results reveal that doping the 1aFe with X atoms, at the equidistant point between the two Fe atoms, reduces the magnetic moment of Fe atoms that particularly in the 1aFeX systems with X=F and B, where the magnetic moment approaches zero. The change in the magnitude of magnetic moment of Fe atoms can be explained within the frame work of the concept of the electron-transfer from X atoms to the $d$ orbitals of the Fe atoms. On the other hand, X doping significantly reduce the exchange couplings along the diagonal and the edge directions. The decrease in the strength of the exchange couplings is attributed to the decrease in the delocalized states in Fe atoms near the Fermi level. Due to the fact that the strength of the RKKY coupling observed in the 1aFe system depends on the delocalized states of Fe atoms near the Fermi level. Moreover, the increase in lattice constant in the 2aFe system enhances the magnetic moment of Fe atoms. This is attributed to the increase in PDOS for the $t_{2g}$ and $e_g$ orbitals of Fe atoms, along with the greater growth in the DOS of spin-up electrons compared to the spin-down electrons. Additionally, the $t_{2g}$ and $e_g$ orbitals of Fe atoms are localized and exhibit no overlap with one another, in contrast to the 1aFe system.

Another effect of the increasing lattice constant is the suppression of exchange couplings along the diagonal and the edge directions. The localization of $t_{2g}$ and $e_g$ states of Fe atoms near the Fermi level in the 2aFe system is an indication that the RKKY mechanism is not dominant in this system. Instead, the dominant mechanism is direct exchange coupling due to the overlap between the mentioned atomic orbitals. However, due to the increased Fe-Fe distance the overlap between atomic orbitals has decreased significantly, resulting in a considerable decrease in exchange couplings along the diagonal and the edge directions.

Furthermore, our results demonstrate that doping the 2$a$Fe system with X atoms different separation lengths from Fe-X, dramatically enhances the exchange coupling specially $J_2$ coupling compared to that of the pristine 2$a$Fe system. Such an increase is due to the formation of an indirect superexchange coupling between Fe atoms, in consequence of the overlapping of $p$ orbitals of X atoms with $d$ orbitals of the Fe atoms. The detailed investigations of the $J_2$ reveal that each 2$a$FeXL$_i$ systems exhibit a maximum coupling strength ($J_2^{max}$) at a specific Fe-X distance. The sensitivity of this coupling to slight variations in distance is highlighted by scanning the Fe-X separation lengths associated with $J_2^{max}$. Further analyses of the $J_2$ exchange couplings confirm that each of the 2$a$FeXL$_s$ systems have distinct distance for the related $J_2^{max}$. We have compared the obtained $J_2^{max}$ values of the doped systems. It reveals that doped systems with hydrogen and fluorine atoms exhibit the highest and lowest $J_2^{max}$ values, respectively among the X atoms.

On other hand, the intensity of the integrated area of unoccupied $d$ of doped 2$a$FeXL$_s$ systems with $J_2^{max}$ in different energy intervals revealed that doped 2$a$Fe systems with X=H, N, C, and O doped atoms as bridges in the exchange mechanism have a similar order. While the delocalization of the integrated area for the hydrogen doped system is greater than that of other systems. This is the reason for the higher $J_2^{max}$ exchange coupling value for the hydrogen doped 2$a$Fe system in a specific distance compared to other doped 2$a$Fe systems. This is an indication that the varying chemical properties and nature of X atoms acting as a bridge in the exchange coupling through superexchange.

We emphasize that our findings reveal the significant role of the X atom as a bridge influencing exchange coupling strength through superexchange. This work also sheds light on the complicated interplay between structure and magnetism in magnetic materials, particularly examining the effects of different chemical environments on exchange couplings. This information is of great importance to experimentalists, establishing a basis for probing different chemical environments in synthesized materials. By applying these insights, researchers could potentially discover new materials that exhibit either enhanced or unique magnetic properties, contributing to the development of high-performance magnetic materials.

## ASSOCIATED CONTENT

Supporting Information

Scanning the Fe-X separation length ($L_s$) that illustrates the maximum exchange coupling $J_2$, is presented in this file.

## ACKNOWLEDGMENTS

F.N. and J. A. are grateful to the *Institute for Advanced Studies in Basic Sciences* for financial support through research Grant No. G2023IASBS32604 and No. GIASBS12969, respectively.

**Table 1** The different Fe-X separation in $2a\text{FeXL}_i$ systems.

| $L_1(\text{Å})$ | $L_2(\text{Å})$ | $L_3(\text{Å})$ | $L_4(\text{Å})$ | $L_5(\text{Å})$ | $L_6(\text{Å})$ | $L_m(\text{Å})$ |
|---|---|---|---|---|---|---|
| 1.655 | 1.763 | 1.832 | 1.843 | 2.130 | 2.195 | 2.860 |

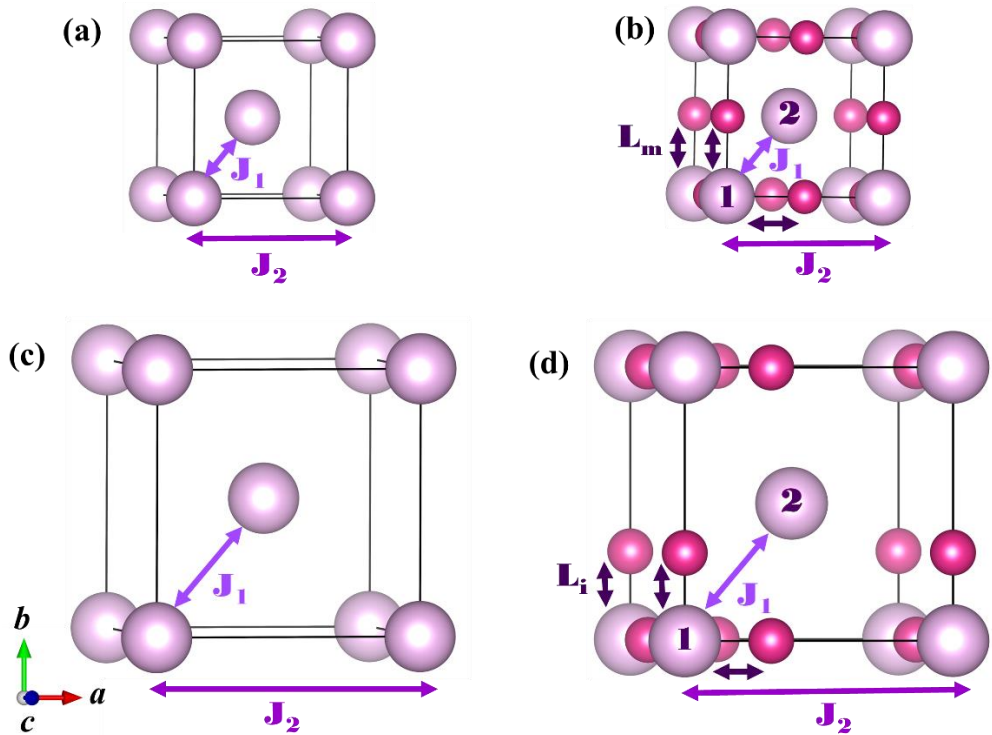

**Fig. 1** Geometry of the studied systems (a) bcc-Fe bulk with the lattice constant $a$ (1$a$Fe), (b) the bcc-Fe bulk with the lattice constant $a$ doped with X atoms (1$a$FeX) (c) the bcc-Fe bulk with the lattice constant $2a$ (2$a$Fe), and (d) the bcc-Fe bulk with the lattice constant $2a$ doped with X atoms (2$a$FeXL$_i$). The gray, and magenta balls stand for Fe and X atoms, respectively. J$_1$ and J$_2$ are exchange couplings between Fe atoms along the diagonal and the edge directions of the cube, respectively. In doped bcc-Fe, the Fe atoms located at the vertices are denoted as Fe1, and the Fe atoms located at the center of the cube are denoted as Fe2.

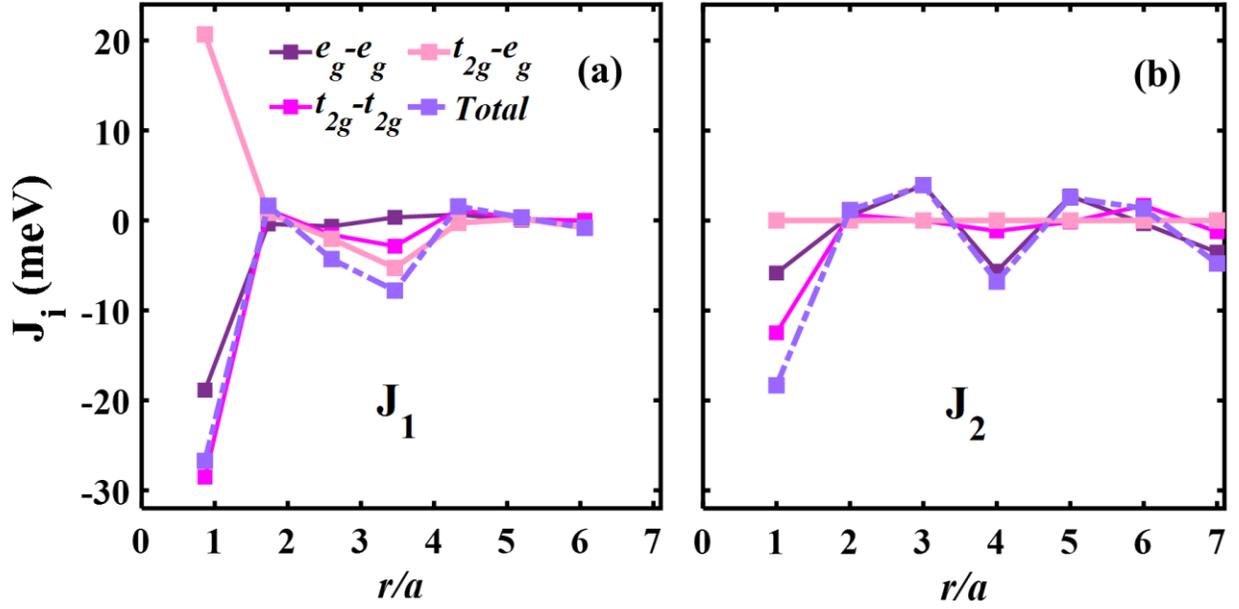

**Fig. 2** Different orbital contributions to the exchange couplings as a function of Fe-Fe interatomic distance (*r/a*) along (a) the diagonal and (b) the edge directions for the 1*a*Fe system.

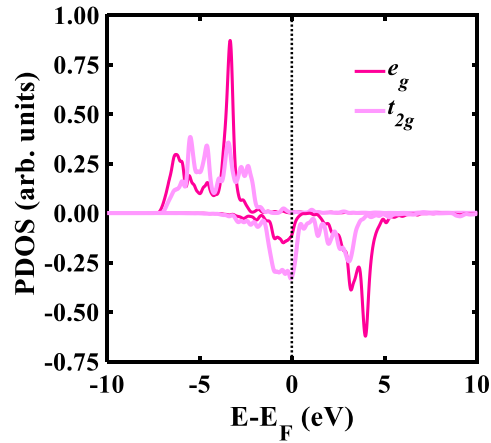

**Fig. 3** The spin-polarized projected density of states (PDOS) of the $e_g$ and $t_{2g}$ orbitals for the 1*a*Fe system.

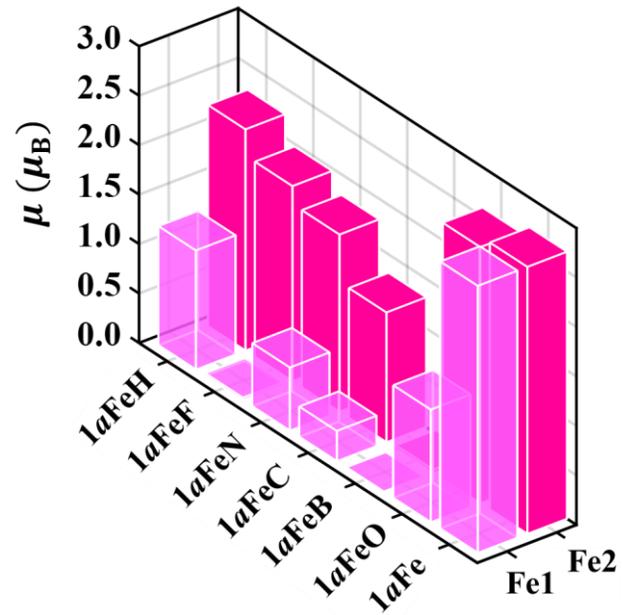

**Fig. 4** The magnetic moment ($\mu$) of the Fe atom located at the vertices and the center of the cube of the 1$a$FeX systems with X being the H, F, N, C, B, and O elements. The magnetic moment of the Fe atom in 1$a$Fe system is also illustrated for comparison. In this case there is no differences between Fe1 and Fe2 atoms, and the magnetic moment of the Fe atoms is 2.7$\mu_B$.

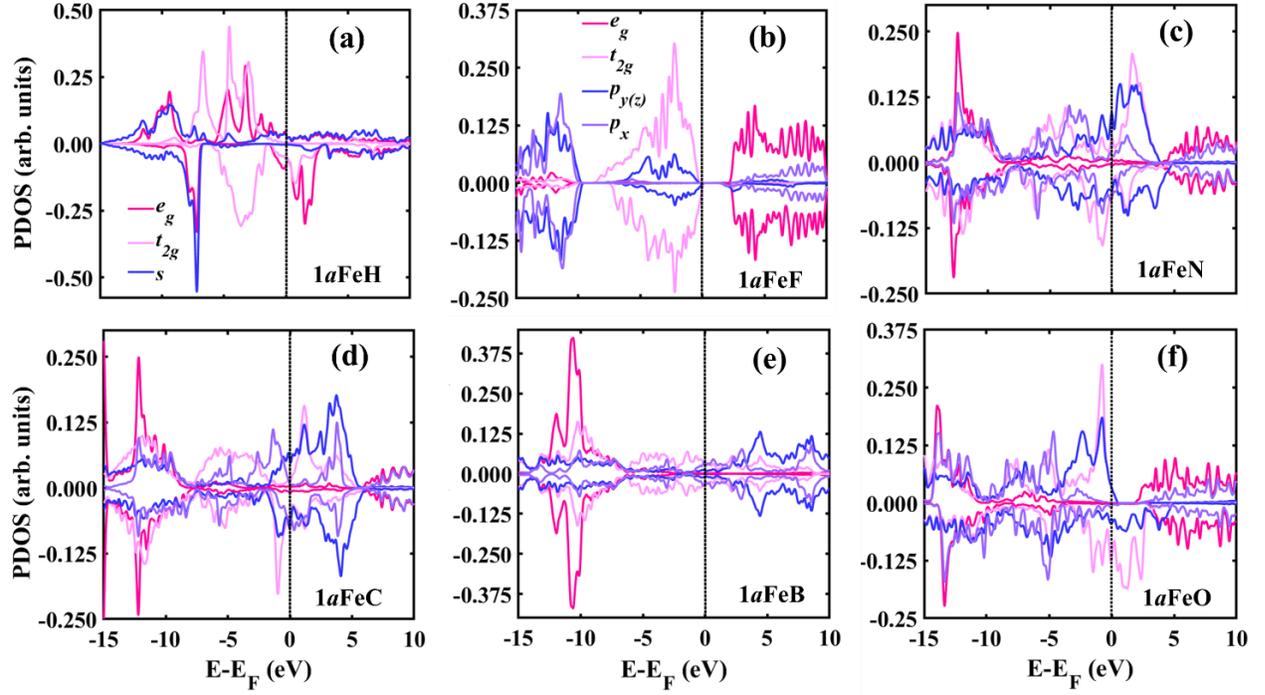

**Fig. 5** The spin-polarized projected density of states (PDOS) for different orbitals of the 1aFeX, systems. Here, the DOS for the $e_g$ and $t_{2g}$ orbitals of the Fe1 atom, the DOS for $p_{x,y,z}$ orbitals of the X atom in 1aFeX system with X=F, N, C, B, O atoms, and the DOS for the $s$ orbital of the H atom in 1aFeH system are illustrated.

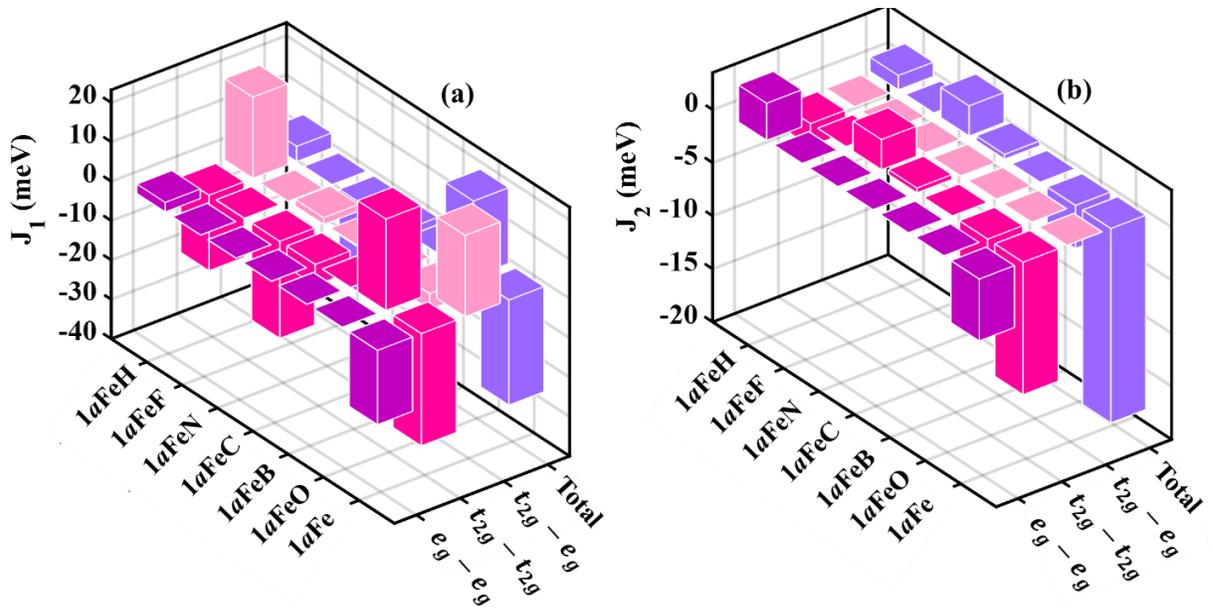

**Fig. 6** The $J_1$ and $J_2$ exchange couplings in the different $1a$FeX systems where X= H, F, N, C, B, and O. Different orbital contributions to the $J_1$ and $J_2$ exchange couplings are also illustrated.

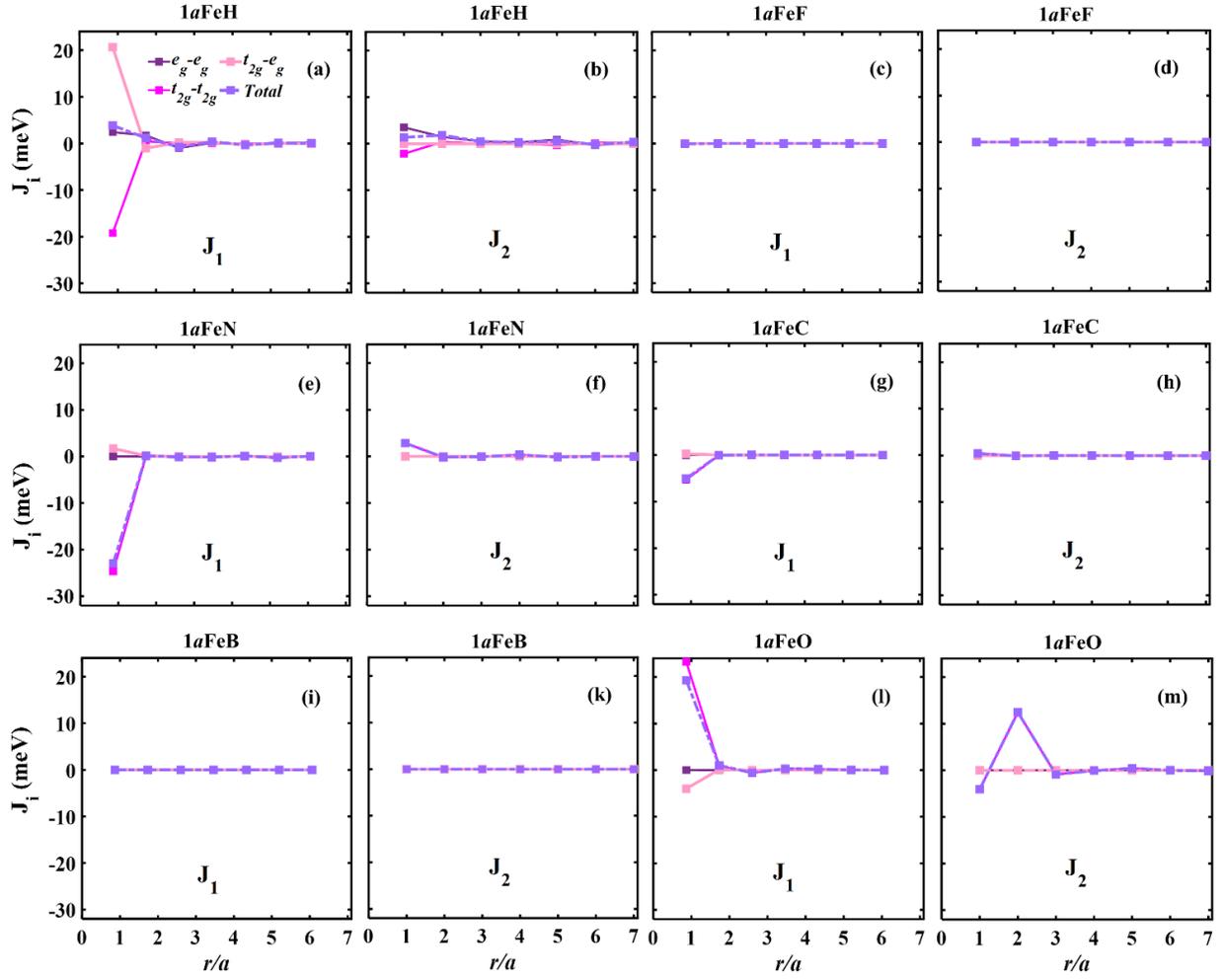

**Fig. 7** Different orbital contributions to exchange couplings as a function of Fe-Fe interatomic distance (*r/a*) for the diagonal and the edge directions of different 1*a*FeX systems.

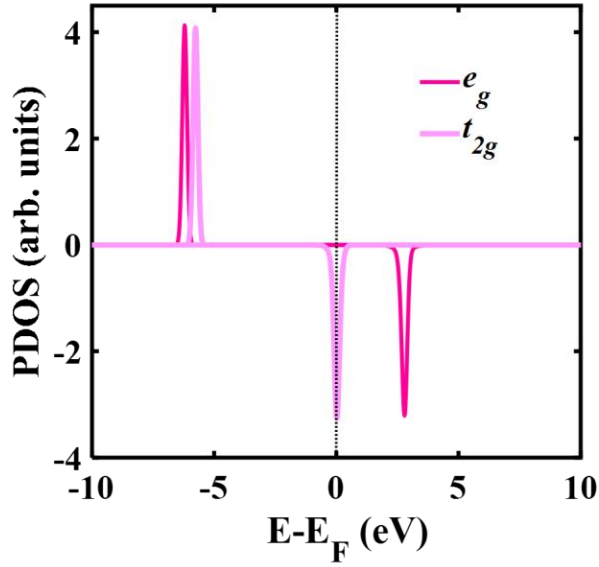

**Fig. 8** The spin-polarized projected density of states (PDOS) for the 2$a$Fe system.

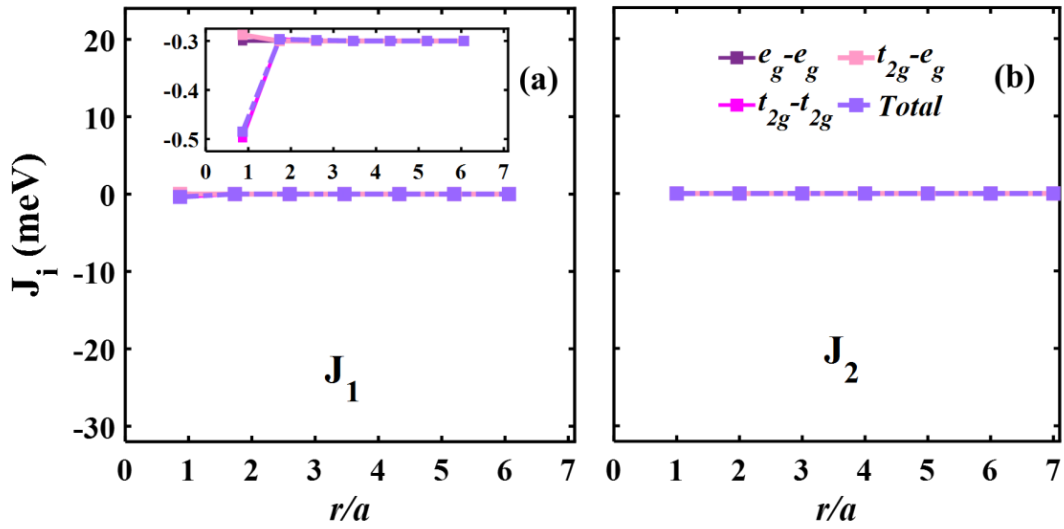

**Fig. 9** Different orbital contributions to the exchange couplings as a function of Fe-Fe separation distance ($r/a$) along the diagonal and the edge directions for 2$a$Fe system.

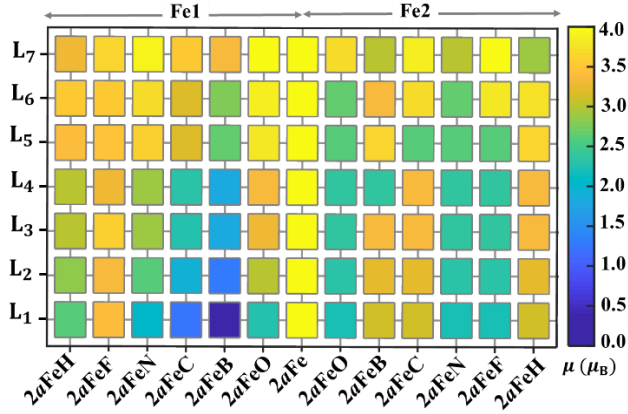

**Fig. 10** The magnetic moment ($\mu$) of the Fe1 and Fe2 atoms located respectively at the vertices and center of the cube of the $2a\text{FeXL}_i$ systems. $L_1,…, L_6$ are the Fe-X separations presented in Table 1. Here, $L_7$ is the $L_m$ where the X atoms are located right at the middle of two Fe1 atoms. The magnetic moment of the Fe atom in $2a$Fe system is also illustrated for comparison. In this case there is no differences between Fe1 and Fe2 atoms, and the magnetic moment of the Fe atoms is $4\mu_B$.

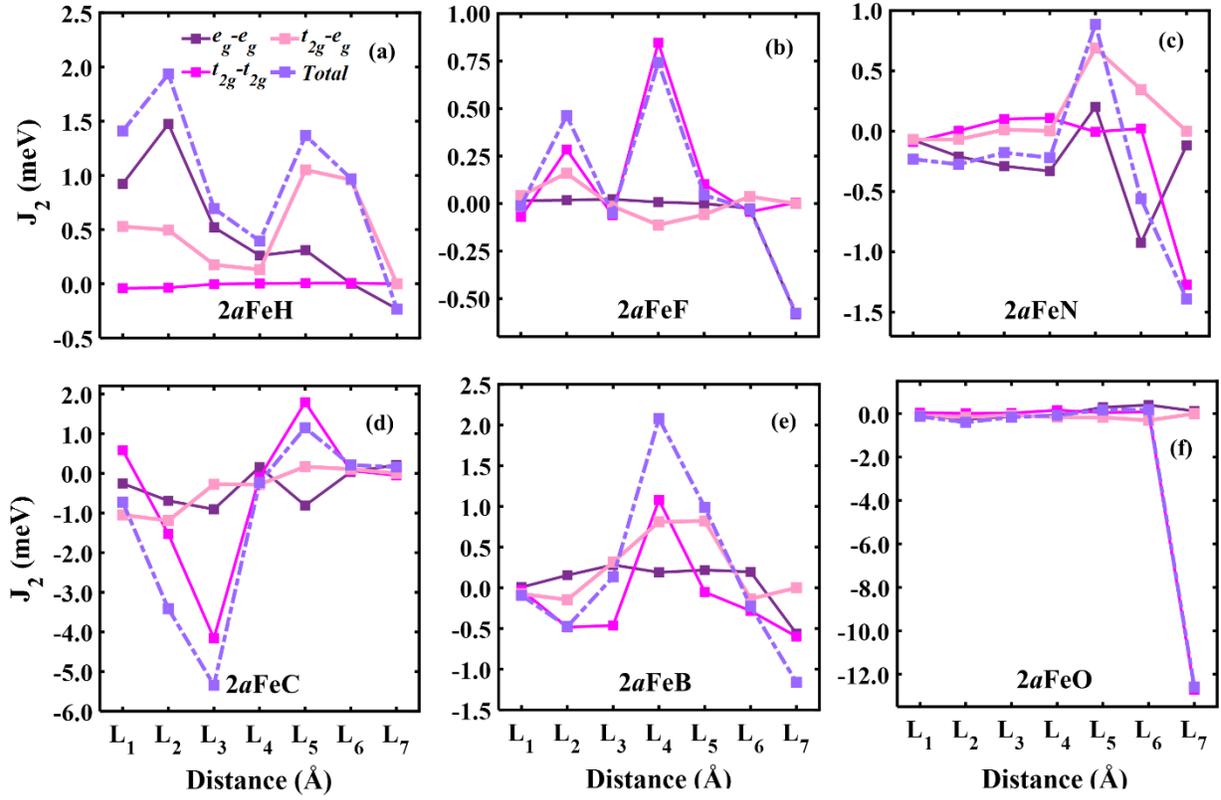

**Fig. 11** The exchange coupling $J_2$ as a function of Fe-X separation distance ($L_i$) in $2a$FeX$L_i$ systems. Here, $L_7$ is the $L_m$, where the X atoms are located right at the middle of two Fe1 atoms. Different orbital contributions are also illustrated.

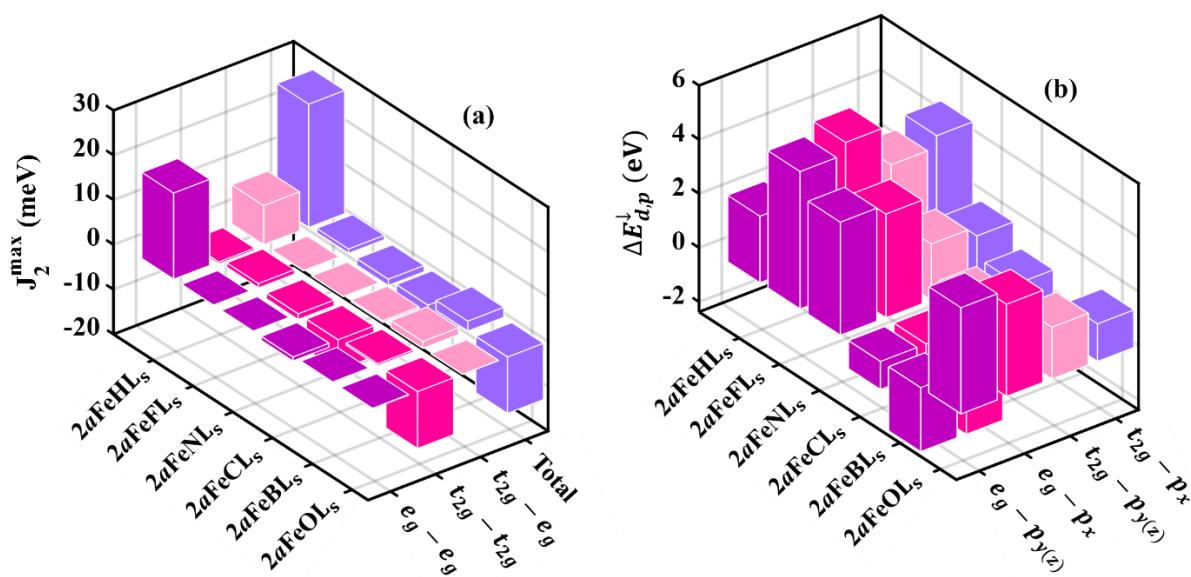

**Fig. 12** (a) The different orbital contributions to $J_2^{max}$. (b) The energy difference between the involved $d$ and $p$ orbitals in each $2a$FeXL$_s$ systems with $J_2^{max}$. L$_s$ is the scanned range. The distance where the $J_2^{max}$ is observed across the entire scanned range is distinct for each $2a$FeXL$_s$ system and is specifically highlighted in Table S1.

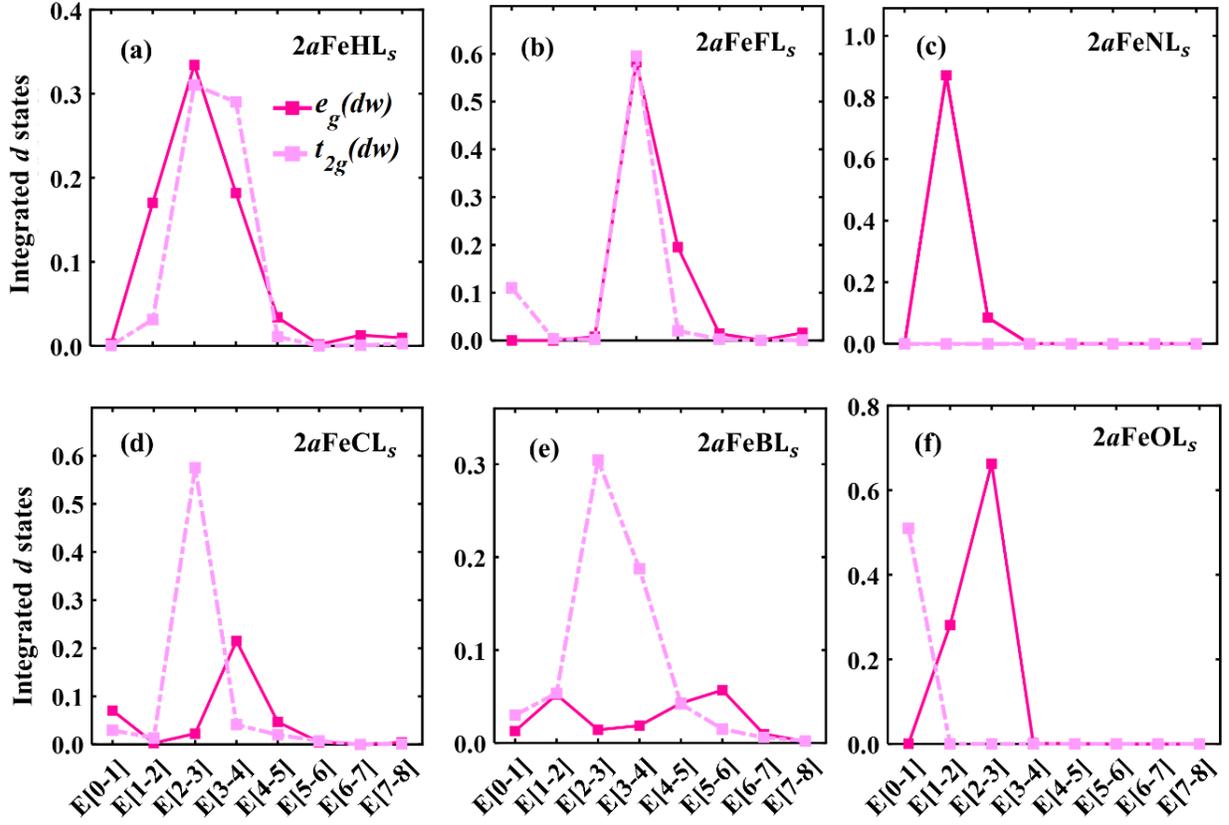

**Fig. 13** The integrated area of the unoccupied $e_g$ and $t_{2g}$ states (spin-down $d$ states) vs $E - E_F$ in different energy intervals of $2a\text{FeXL}_s$ systems with $J_2^{max}$.

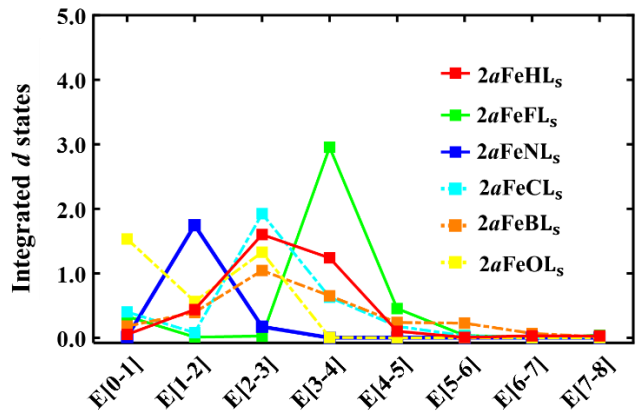

**Fig. 14** The integrated area of the unoccupied $d$ ($e_g + t_{2g}$) states (spin-down $d$ states) vs $E - E_F$ in different energy intervals for $2a\text{FeXL}_s$ systems with $J_2^{max}$.